# ON THE ADHESION OF PARTICLES TO A CELL LAYER UNDER FLOW

*F. Gentile [a], A. Granaldi [b], M. Ferrari [c], P. Decuzzi [a,b,c] **

[a] Center of Bio-/Nanotechnology and -/Engineering for Medicine (BioNEM), University of Magna Graecia, Catanzaro - ITALY
[b] Center of Excellence in Computational Mechanics (CEMEC), Politecnico di Bari, Bari - ITALY
[c] The Univeristy of Texas Health Science Center at Houston, Houston, Texas - USA

**ABSTRACT**

The non-specific adhesion of spherical particles to a cell substrate is analyzed in a parallel plate flow chamber, addressing the effect of the particle size. Differently from other experiments, the total volume of the injected particles has been fixed, rather than the total number of particles, as the diameter *d* of the particles is changed from 500 nm up to 10 μm. From the analysis of the experimental data, simple and instructive scaling adhesion laws have been derived showing that (i) the number of particles adherent to the cell layer per unit surface decreases with the size of the particle as $d^{-1.7}$; and consequently (ii) the volume of the particles adherent per unit surface increases with the size of the particles as $d^{1.3}$. These results are of importance in the 'rational design' of nanoparticles for drug delivery and biomedical imaging.

## 1. INTRODUCTION

Advances in nanosciences and micro/nanofabrication techniques have led to novel strategies and devices for drug delivery and diagnosis in biomedical applications [1,2]. In particular, biomimetic artificial carriers are emerging as powerful tools for cancer and heart diseases treatment and molecular imaging. These are particulates carrying therapeutic and/or imaging agents administered at the systemic level and delivered towards a specific biological target (diseased cell or microenvironment). The surface of these particulates is generally coated by ligand molecules that can recognize countermolecules (receptors) expressed at the biological target. The power of intravascularly injectable particles over free molecules administration lies in their multifunctionality and engineerability [3]. The use of such particles as intravascular delivery vectors affords substantial advantages including (i) the specific biomolecular targeting through one or more conjugated antibodies or other recognition molecules; (ii) the ability to carry one or more therapeutic agents; (iii) the signal amplification for imaging through co-encapsulated contrast agents; (iv) the tailoring of the physico-chemical and geometrical properties favoring avoidance of the biological and physiological barriers and improve the recognition of the target cells or microenvironment.

A broad spectrum of particulates have been presented in the literature which can be used both in medical therapy and imaging. These have different sizes, ranging from few tens of nanometers (dendrimers [4]) to hundreds of nanometers (polymer particles and liposome [5,6]) up to few microns (silicon and silica based particles [7]); different shapes from the classical spherical to spheroidal [8] and even more complex shapes [9]; and with different compositions and surface chemico-physical properties. Differently from freely administered drugs, these particles can be engineered [10,11] and it is of fundamental importance to understand if, in the great variety of shapes, sizes and compositions, there is any 'optimal particle' with the largest selectivity and delivery efficiency.

In this work, the adhesive performance of the particles with a size ranging between 500 nm and 10 μm is assessed in terms of the number of particles adhering non-specifically to a confluent layer of endothelial cells under fixed hydrodynamic conditions. Goetz and colleagues [12] have already analyzed the effect of the size of spherical particles on their adhesive performance. However, since they were interested in analyzing the interaction between leukocytes and endothelial cells, employed 5−, 10−, 15−, and 20 − μm diameter microspheres. Also, the experimental analysis was carried out using a biological substrate consisting of a glass slide covered with P-selectin molecules, whereas in this work the biological substrate is made up of a confluent layer of endothelial cells. Nonetheless, they showed that the adhesive strength of the particles, expressed in terms of the critical shear stress and rolling velocity, is significantly influenced by their diameter.

## 2. MATERIALS AND METHODS

A Glycotech (Rockville, MD) parallel plate flow chamber has been used as described in [13]. The flow chamber is connected to a Harvard Apparatus syringe pump through the inlet hole and a silastic tubing; and to a reservoir through the outlet hole and a silastic tubing. The channel within the flow chamber is defined by a silicon rubber gasket sitting between the flow deck and the microscope 35-mm dish. The gasket used in the present experiments has a thickness of 0.01 in and a width of 1 cm. The volumetric flow rate Q has been fixed equal to 50 μl/min for all the experiments. After assembling the flow chamber on the stage of an inverted microscope, a region at the center of the channel is chosen with a sufficiently large number of confluent cells. This is the region of interest (ROI). The microscope objective is focused on this region and after a short rinse, the flow with the particles enters the chamber.

Using the fluorescent module of the microscope, the flow of the particles in the ROI is recorded during the whole experiments. Each acquired frame is then analyzed in MatLab® (Matlab6.5) using a self-developed imaging analysis code. The number of particles adherent to the whole substrate $n_{tot}$ within the ROI and the number of particles adherent to the sole cells $n_c$ within the ROI are measured as a function of the frame number, which is easily converted in time by knowing the acquisition rate.

Human umbilical vein endothelial cells (HUVECs), purchased from Cambrex, Inc. (East Rutherford, NJ), were plated on a borosilicate glass with a 0.2 mg/cm² substratum of type A gelatine (Sigma-Aldrich Corporation, MO).

Fluoresbrite® Microspheres from Polysciences were used.

## 3. RESULTS

The number of adherent particles within the region of interest (ROI) has been monitored as a function of the size during the whole experiment. Figure1 shows, as an example, the variation with time of the particles $n_c$ adhering solely to the cells (lower curve) and the total number $n_{tot}$ of adherent particles within the ROI (upper curve), for the 1 μm, 750 and 500 nm particles. An image of the fluorescent particles adhering to the culture dish and to the cells is shown in Figure2, for the 1 μm particle. From the analysis of the experimental data, the number of particles $n_{tot}$ and $n_c$ at the end of each experiment can be plotted as a function of the particle diameter *d* (Figure3). As the diameter of the particle reduces the number of adherent particles, both $n_{tot}$ and $n_c$, increases; and a non linear regression of the experimental data leads to the two following approximate expressions

$$n_{tot} = 627.96 d^{-1.696} \quad (1)$$

$$n_c = 164.754 d^{-1.143} \quad (2)$$

Differently, in Figure4, the number of particles adherent per unit surface is shown as a function of their diameter: the total number of adherent particles $n_{tot}$ is normalized by the area of the ROI, whereas the number of particles adherent to the sole cells $n_c$ is normalized by the area occupied by the cells within the ROI, which depends on the level of cell confluency. In a double logarithm diagram, the surface densities of the total number of particles $\tilde{n}_{tot}$ and of the particles adherent on the cells $\tilde{n}_c$ are well aligned along two straight and almost parallel lines (Figure4), and the non linear regression of the experimental data leads to

$$\tilde{n}_{tot} = 1116.39 d^{-1.696} \quad (3)$$

$$\tilde{n}_c = 2784.34 d^{-1.746} \quad (4)$$

## 4. DISCUSSION AND CONCLUSIONS

The non-specific adhesion under flow of spherical particles onto a confluent layer of endothelial cells is analyzed in-vitro using a parallel plate flow chamber. Differently from other experimental analysis, the total number of particles injected in the flow chamber has been changed to keep fixed the total volume of the particles with their size. Since the focus is on drug delivery and biomedical imaging, by keeping fixed the total volume of the particles, the total volume of therapeutic or imaging agents delivered to the cell layer are fixed in the analysis. The experimental results, which have been derived under fixed hydrodynamic conditions, lead to conclude that (i) the surface density of particles adhering to the cells $\tilde{n}_c$ decreases with the diameter *d* following the scaling law $\approx d^{-1.7}$; (ii) the volume per unit surface of the particles adhering to the cells $\tilde{V}(=\pi \tilde{n}_c d^3/6)$ increases with the diameter *d* following the scaling law $\approx d^{+1.3}$. These scaling laws have been also justified by theoretical reasonings, as described in details in [14].

Since in drug delivery, it is desirable to increase the amount of drug that can be released per unit surface on a cell layer, the above results would suggest to use the largest possible particles in drug delivery. Notice however, that the following suggestion is derived by experimental observations of non-specific adhesion at sufficiently small shear stresses at the wall. In specific adhesion, Decuzzi and Ferrari [15] have already shown that there is an optimal particle size corresponding to the largest strength of adhesion.

Differently in biomedical imaging, it is desirable to increase the number of particles adherent per unit surface to have the largest possible resolution of the cell layer. As a consequence, the above results would suggest to use the smallest possible particles in biomedical imaging.

* **Corresponding Author**:

Paolo Decuzzi, PhD p.decuzzi@poliba.it

Associate Professor of Mechanical and Biomedical Engineering

Center of Bio-/Nanotechnology and -/Engineering for Medicine (BioNEM); University of Magna Graecia, Viale Europa - Loc. Germaneto; 88100 Catanzaro, ITALY
and Center of Excellence in Computational Mechanics (CEMEC), Politecnico di Bari, Via Re David 200 Bari, ITALY

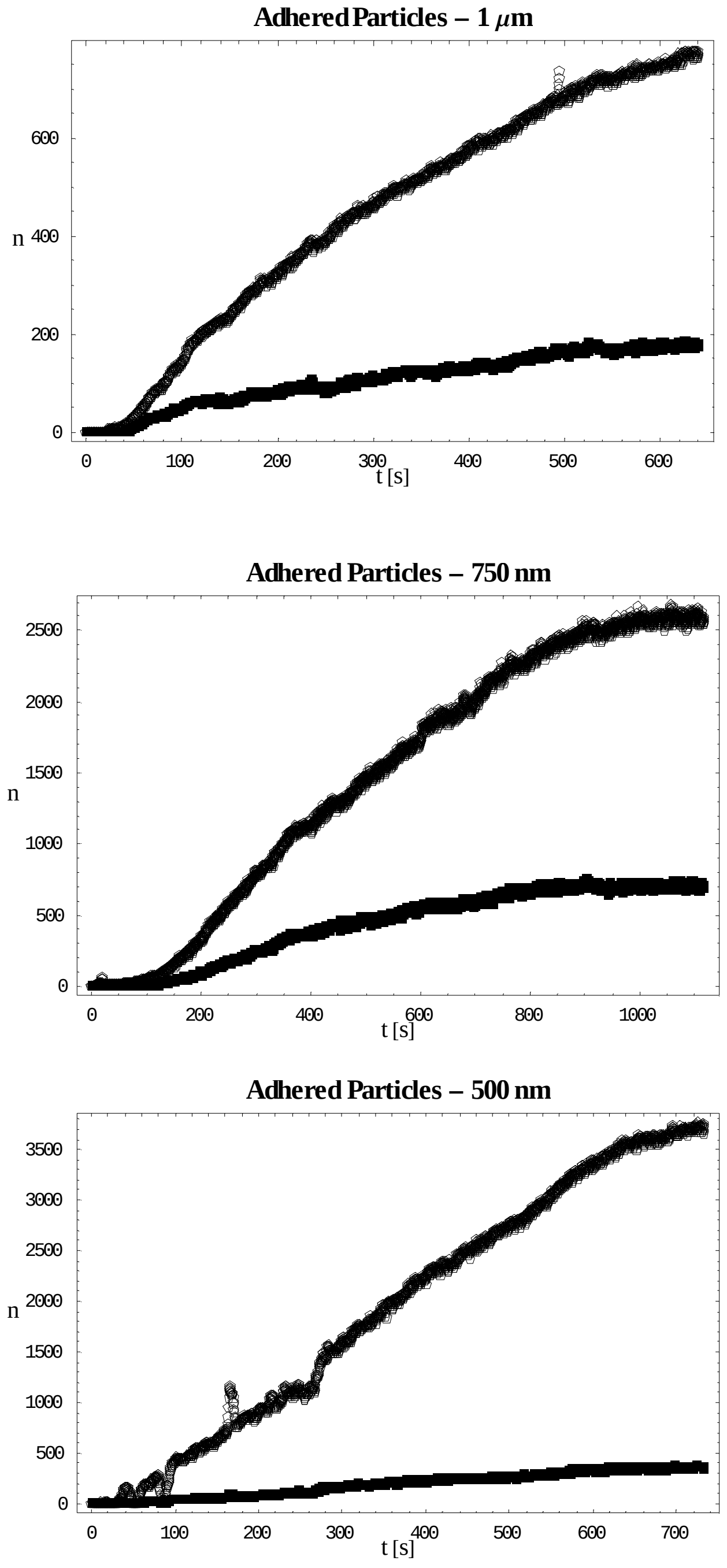

**Figure1.** The variation of the number *n* of adherent particles with time for $d = 1\mu m$, $d = 750\ nm$ and $d = 500\ nm$. The upper curve is for the total number of adherent particles $n_{tot}$, whereas the lower curve is for the particles adhering solely on cells $n_c$.

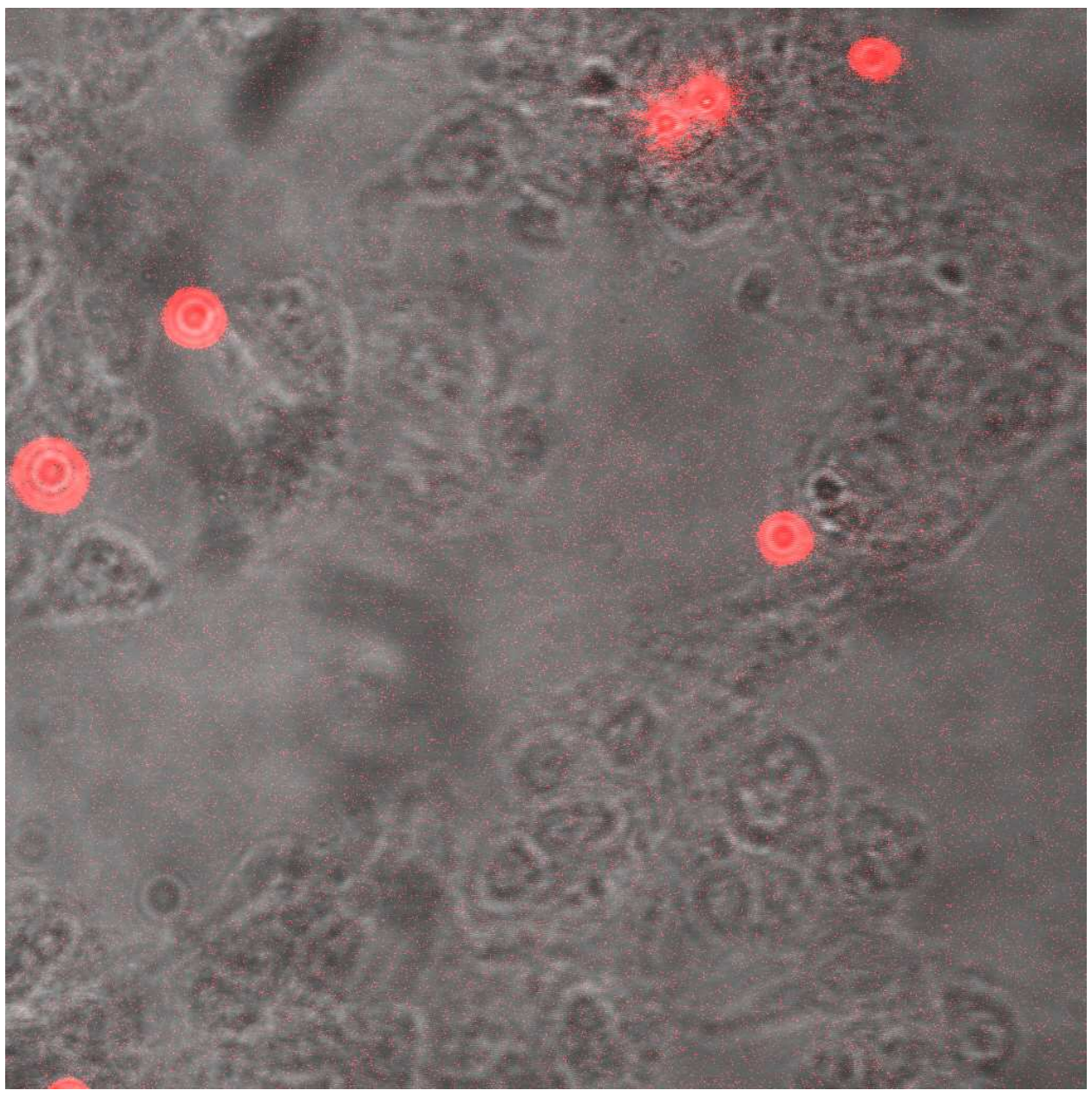

**Figure2.** An image showing red fluorescent *1 µm* - particles adhering to a layer of HUVECs within the region of interest (ROI).

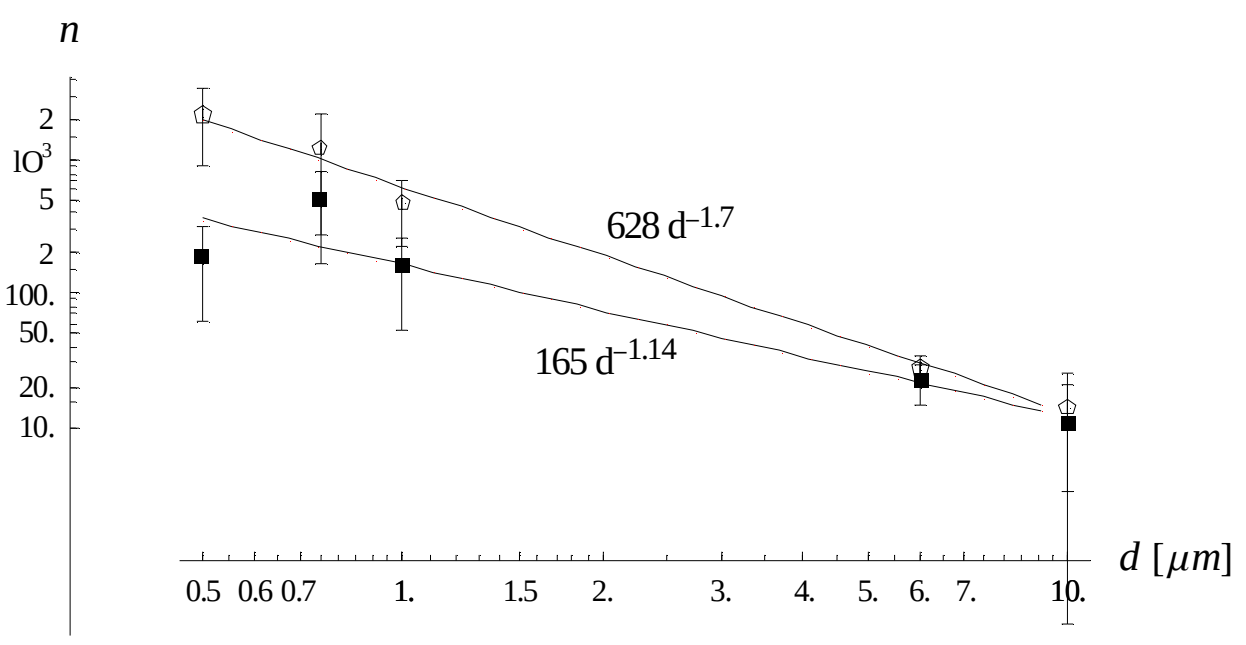

**Figure3.** The variation of the number of particles adhering at the end of the experiments within the ROI, on the sole cells (black boxes) and on the glass dish (white circles) as a function of the particle diameter *d*.

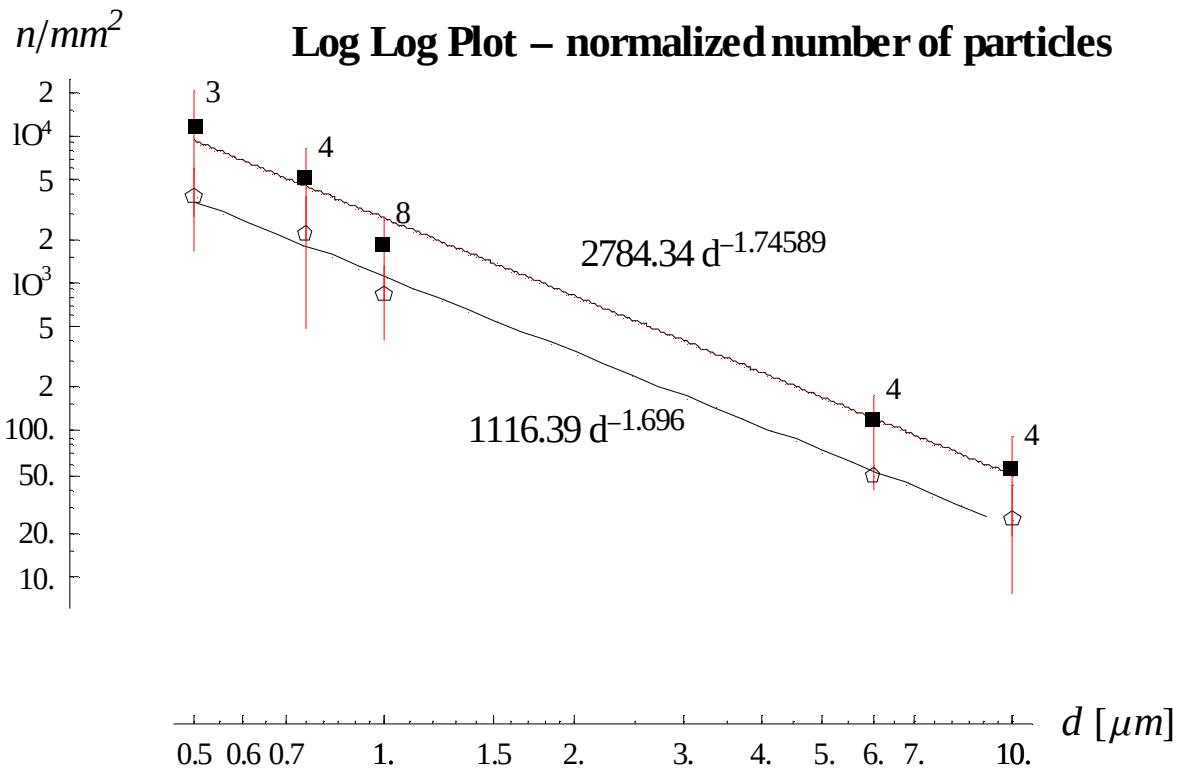

**Figure4.** The variation of the surface density of particles adhering at the end of the experiment within the ROI, on the sole cells (black boxes) and on the glass dish (white circles) as a function of the particle diameter *d*.